\documentclass[equations,12pt]{article}
\usepackage{amsfonts}
\usepackage{a4,tikz}
\usepackage{geometry}
\usepackage{color} 
\usepackage{subfigure}
\makeatletter
\newcommand\ackname{Acknowledgements}
\if@titlepage
  \newenvironment{acknowledgements}{%
      \titlepage
      \null\vfil
      \@beginparpenalty\@lowpenalty
      \begin{center}%
        \bfseries \ackname
        \@endparpenalty\@M1
      \end{center}}%
     {\par\vfil\null\endtitlepage}
\else
  \newenvironment{acknowledgements}{%
      \if@twocolumn
        \section*{\abstractname}%
      \else
        \small
        \begin{center}%
          {\bfseries \ackname\vspace{-.5em}\vspace{\z@}}%
        \end{center}%
        \quotation
      \fi}
      {\if@twocolumn\else\endquotation\fi}
\fi
\makeatother
\usepackage{centernot}
\usepackage{mathtools}
\usepackage{ stmaryrd }
\usepackage{amsmath}
\usepackage{amssymb}
\usepackage{amsthm}
\usepackage{mathrsfs}
\usepackage{eucal}
 \usepackage[usenames,dvipsnames]{pstricks}
 \usepackage{epsfig}
 \usepackage{pst-grad} 
 \usepackage{pst-plot} 
\renewcommand{\theequation}{\arabic{equation}}
\usepackage{amssymb,amsmath}
\usepackage{color}
\usepackage{accents}
\usepackage{texdraw}
\usepackage{eucal}
\usepackage{epic,epsfig}
\usepackage{graphicx}

\theoremstyle{definition}

\numberwithin{equation}{section}
\DeclareMathAccent{\wtilde}{\mathord}{largesymbols}{"65}
\DeclareMathAccent{\what}{\mathord}{largesymbols}{"62}

\def\m@th{\mathsurround=0pt}
\mathchardef\bracell="0365
\def\upbrall{$\m@th\bracell$}
\def\undertilde#1{\mathop{\vtop{\ialign{##\crcr
    $\hfil\displaystyle{#1}\hfil$\crcr
     \noalign
     {\kern1.5pt\nointerlineskip}
     \upbrall\crcr\noalign{\kern1pt
   }}}}\limits}
\def\m@th{\mathsurround=0pt}
\mathchardef\bracell="0365
\def\upbrall{$\m@th\bracell$}
\def\underhat#1{\mathop{\vtop{\ialign{##\crcr
    $\hfil\displaystyle{#1}\hfil$\crcr
     \noalign
     {\kern1.5pt\nointerlineskip}
     \upbrall\crcr\noalign{\kern1pt
   }}}}\limits}
\usepackage{pgf}
\usepackage{tikz}
\usepackage{verbatim}
\usepackage[utf8]{inputenc}
\usetikzlibrary{arrows,automata}
\usetikzlibrary{positioning}

\def\theequation{\arabic{section}.\arabic{equation}}

\newcommand{\wh}{\widehat}
\newcommand{\wt}{\widetilde}

\def\hypotilde#1#2{\vrule depth #1 pt width 0pt{\smash{{\mathop{#2}
\limits_{\displaystyle\widetilde{}}}}}}
\def\hypohat#1#2{\vrule depth #1 pt width 0pt{\smash{{\mathop{#2}
\limits_{\displaystyle\widehat{}}}}}}

\newcommand{\tbu}{\,^{t\!}{\bu}}

\newcommand{\bblu}{\begin{color}{blue}}
\newcommand{\bred}{\begin{color}{red}}
\newcommand{\ecl}{\end{color}}

\newcommand{\bsY}{\boldsymbol{Y}}

\newcommand{\bLam}{\boldsymbol{\Lambda}}

\newcommand{\be}{\begin{equation}}
\newcommand{\ee}{\end{equation}}
\newcommand{\bea}{\begin{eqnarray}}
\newcommand{\eea}{\end{eqnarray}}
\newcommand{\bse}{\begin{subequations}}
\newcommand{\ese}{\end{subequations}}
\newcommand{\nn}{\nonumber}

\newcommand{\ol}{\overline}

\newcommand{\bu}{\boldsymbol{u}}

\begin{document}

\def\theequation{\arabic{section}.\arabic{equation}}

\newtheorem{thm}{Theorem}[section]
\newtheorem{lem}{Lemma}[section]
\newtheorem{defn}{Definition}[section]
\newtheorem{ex}{Example}[section]
\newtheorem{rem}{}
\newtheorem{criteria}{Criteria}[section]
\newcommand{\ra}{\rangle}
\newcommand{\la}{\langle}

\title{\textbf{On the Lagrangian structure of Calogero's Goldfish model}}
\author{\\\\Umpon Jairuk$^\dagger$, Sikarin Yoo-Kong$^{\dagger,* } $ and Monsit Tanasittikosol$^{\dagger} $ \\
\small $^\dagger $\emph{Theoretical and Computational Physics (TCP) Group, Department of Physics,}\\ 
\small \emph{Faculty of Science, King Mongkut's University of Technology Thonburi,}\\
\small \emph{Thailand, 10140.}\\
\small $^* $ \emph{Ratchaburi Campus, King Mongkut's University of Technology Thonburi,}\\
\small \emph{Thailand, 70510.}\\
}
\maketitle


\abstract
The discrete-time rational Calogero's goldfish system is obtained from the Ansatz Lax pair. The discrete-time Lagrangians of the system possess the discrete-time 1-form structure as those in the discrete-time Calogero-Moser system and discrete-time Ruijsenaars-Schneider system. Performing two steps of continuum limits, we obtain Lagrangian hierarchy for the system. Expectingly, the continuous-time Lagrange 1-form structure of the system holds. Furthermore, the connection to the lattice KP systems is also established. 

\section{Introduction}\label{intro}
\setcounter{equation}{0}

The multi-dimensional consistency plays a very important role for the notion on integrability of the discrete systems. In the nutshell, for any $D$-dimensional discrete system, we find that the system in higher dimensions (spaces and times) can be compatibly constructed from the subsystems in lower dimensions (spaces and times). The number of dimensions $D$ can be set to be infinity which in this case we could have an infinite set of compatible subsystems. 

According to the least action principle in classical mechanics, the action of the system is stationary for the classical path on space constituted from dependent variable(s) and independent variable(s). Then we may ask what is the analogue for the least action principle for the systems satisfying the multi-dimensional consistency. Imagine that not only we consider the path in the subspace constituted from dependent variable(s) and independent variable(s), but also the subspace of independent variable(s). Recently, there has been a  theory, called the Lagrangian multiform theory, initiated by Sarah Lobb and Frank Nijhoff \cite{SF1,SF2,SF3}, which tried to address the above question, explicitly for the case $D=2$ \cite{SF1,SF2} and $D=3$ \cite{SF3}. The key idea of this theory is that the action of the systems is invariant under the variation on the independent variables resulting in the feature relation called the closure relation which can be considered to be representation of the multi-dimensional consistency in Lagrangians aspect. For the case $D=1$, the concrete model called the rational Calogero-Moser system which is the many-body system in one dimension with a long range interaction \cite{Original CM1,Original CM2}, was studied in both discrete time and continuous time \cite{Sikarin1}. In this case, the Lagrangians satisfy the 1-form structure. Then soon after the rational Ruijsenaars-Schneider system (considered to be relativistic version of the Calogero-Moser system) was also studied in the full detail of its Lagrangian structure\cite{Sikarin2}. In the case of one dimensional many-body system with nearest neighbour interaction called the Toda-typed system was also studied in the discrete level \cite{Toda,Toda2}.

In this paper, we consider the system called the rational Calogero's goldfish \cite{Calogero goldfish} system in order to complete the big picture of the Lagrangian 1-form theory for the integrable one-dimensional many-body systems with long range interaction. Interestingly, the Calogero's goldfish system can be reduced from the Ruijsenaars-Schneider system by setting the relativistic parameter to be infinity (for relativistic parameter approaches to zero the system will go to the Calogero-Moser system). 
The organisation of the paper is the following. In section \ref{exactsolution}, the full details at the level of discrete-time of the system will be carried out. The variation of discete-time action constituted from the discrete curves will be computed resulting in the discrete-time Euler-Lagrange as well as the closure relation. In section \ref{skewlimit}, the first continuum limit called the skew limit will be computed leaving the system in semi-discrete level. In section \ref{fullLIMIT}, the second continuum limit will be performed to get rid of the remaining discrete variable resulting in the system in fully continuous level. In section \ref{conKP}, the connection to the lattice KP systems is established through the structure of the exact solution of the system. In the last section 6, the summary of the paper will be given.

\section{The discrete-time Goldfish system and commuting flows}\label{exactsolution}
\setcounter{equation}{0} 
In this section, we will construct the discrete time Calogero's goldfish system. We first consider the system of linear equations

\begin{subequations}\label{LM} 
\begin{eqnarray}
\boldsymbol L_\kappa\boldsymbol \phi&=&\zeta \boldsymbol \phi\ ,\label{LM12}\\
\boldsymbol M_\kappa\boldsymbol \phi&=&\wt{\boldsymbol \phi}\  ,\label{LM13}\\ 
\boldsymbol N_\kappa\boldsymbol \phi&=&\wh{\boldsymbol \phi}\,,\label{LM14} 
\end{eqnarray}
where $\boldsymbol\phi=\boldsymbol\phi(n,m)$ is a vector function, $\zeta $ is an eigenvalue. Here the variables $(n,m)$ are the discrete-time variables such that $\wt{\boldsymbol \phi}=\boldsymbol\phi(n+1,m)$ and $\wh{\boldsymbol \phi}=\boldsymbol\phi(n,m+1)$.
\\
\\
For the rational case, we take the $\boldsymbol L_\kappa$, $\boldsymbol M_\kappa$ and $\boldsymbol N_\kappa$ in the forms
\begin{eqnarray}\label{LM121}
\boldsymbol L_{\kappa}&=&\frac{hh^T}{\kappa}+\boldsymbol L_0\;,\label{LM1}\\
\boldsymbol M_{\kappa}&=&\frac{\wt{h}h^T}{\kappa}+\boldsymbol M_0\;,\label{aLM2}\\
\boldsymbol N_{\kappa}&=&\frac{\wh{h}h^T}{\kappa}+\boldsymbol N_0\;,\label{aLN2}\;
\end{eqnarray}
and
\begin{eqnarray}
\boldsymbol L_0&=&\sum_{i,j=1}^N{h_ih_j}E_{ij}\;,\label{L0}\\
\boldsymbol M_0&=&\sum_{i,j=1}^N\frac{\wt{h}_ih_j}{\wt{x}_i-x_j}E_{ij}\;,\label{LM2}\\
\boldsymbol N_0&=&\sum_{i,j=1}^N\frac{\wh{h}_ih_j}{\wh{x}_i-x_j}E_{ij}\;.\label{LN2}
\end{eqnarray}
\end{subequations} 
The $x_i$ is the position of the $i^{th}$ particle and $N$ is the number of particles in the system. The $h_i=h_i(n,m)$ are auxiliary variables. Again we define the notions (will be used throughout the text): $x_i=x_i(n,m)$ and
\begin{eqnarray}
\mbox{Forward shift in tilde direction}:\;\;\;\; x_i(n+1,m)&=&\wt{x}_i\;\nn\\
\mbox{Backward shift in tilde direction}:\;\;\;\; x_i(n-1,m)&=&\undertilde{x_i}\;\nn\\
\mbox{Forward shift in hat direction}:\;\;\;\; x_i(n,m+1)&=&\wh{x}_i\;\nn\\
\mbox{Backward shift in hat direction}:\;\;\;\; x_i(n,m-1)&=&\hypohat 0 {x_i}\;\nn
\end{eqnarray}
and the variable $\kappa$ is the additional spectral parameter. The $E_{ij}$ is the matrices with entries $(E_{ij})_{kl}=\delta_{ik}\delta_{jl}$. 
\\
\\
Next we will look at the compatibility of the system of equations \eqref{LM}. 
\\
\\
\textbf{\emph{First discrete flow}}:
The compatibility between \eqref{LM12} and \eqref{LM13} gives 
\begin{subequations}
\begin{eqnarray}\label{coM}
\wt{\boldsymbol L}_{\kappa}\boldsymbol M_{\kappa}&=&\boldsymbol M_{\kappa}\boldsymbol L_{\kappa} \nn\\
\left( \frac{\wt h\wt h^T}{\kappa}+\wt{\boldsymbol L}_0\right)\left( \frac{\wt{h}h^T}{\kappa}+\boldsymbol M_0\right)
&=&\left(\frac{\wt{h}h^T}{\kappa}+\boldsymbol M_0\right)\left( \frac{hh^T}{\kappa}+\boldsymbol L_0\right)\;.
\end{eqnarray}
Considering the coefficient of $1/\kappa^2$, we have
\begin{equation}
\sum_{j=1}^N\wt{h}_j^2=\sum_{j=1}^Nh_j^2\;,\label{conh}
\end{equation}
and the coefficient of the $1/\kappa$ provides
\begin{equation}\label{iden1}
\wt{\boldsymbol L}_{0}\wt h h^T+\wt h\wt h^T\boldsymbol M_{0}=\boldsymbol M_{0}hh^T+\wt h h^T\boldsymbol L_{0}\;.
\end{equation}
For the rest of \eqref{coM}, we obtain
\begin{eqnarray}\label{iden2}
\wt{\boldsymbol L}_{0}\boldsymbol M_{0}=\boldsymbol M_{0}\boldsymbol L_{0}\;.
\end{eqnarray}
The equations \eqref{iden1} and \eqref{iden2} produce the identical set of equations
\begin{equation}\label{idenforh1}
\sum_{j=1}^N
\frac{\wt{h}_j^2}{\wt{x}_j-{x}_l}
=\sum_{j=1}^N
\frac{h_j^2}{\wt{x}_i-x_j}\;,
\end{equation}
for all $i,j=1,2,...,N$. Since both sides of \eqref{idenforh1} depend on different external indices, we can write a coupled system of equations: 
\begin{eqnarray}\label{idenforh2aa}
\sum_{j=1}^N
\frac{\wt{h}_j^2}{\wt{x}_j-{x}_l}&=&-p\;\;\;\;\;, \forall l\;,
\end{eqnarray}
\begin{eqnarray}
\sum_{j=1}^N
\frac{h_j^2}{\wt{x}_i-x_j}&=&-p \;\;\;\;\;,\forall i\;,
\end{eqnarray}
where $p = p(n)$ is independent of particles' indices, but can still be a function of discrete-time variable $n$. 

In order to determine the function $h_i$, we use
the Lagrange interpolation formula. Consider $2N$ noncoinciding complex numbers $x_k$ and $y_k$, where $k=1,2,...,N$. Then the following formula holds true:
\begin{equation}\label{LaInab}
\prod_{k=1}^N\frac{(\xi-x_k)}{(\xi-y_k)}=1+\sum_{k=1}^N\frac{1}{(\xi-y_k)}\frac{\prod_{j=1}^N(y_k-x_j)}{\prod_{j=1,j\neq k}^N(y_k-y_j)}\;.
\end{equation}
As a consequence
\begin{equation}\label{LaIn2}
-1=\sum_{k=1}^N\frac{1}{(x_i-y_k)}\frac{\prod_{j=1}^N(y_k-x_j)}{\prod_{j=1,j\neq k}^N(y_k-y_j)}\;,\;\;\;\;\;\;i=1,...,N\;,
\end{equation}
which is obtained by inserting $\xi=x_i$ into Eq. \eqref{LaInab}.
\\
\\
Using \eqref{LaIn2}, we obtain    
%
\begin{eqnarray}
h_j^2&=&p\frac{\prod_{i=1}^N(x_j-\wt{x}_i)}{\prod_{j\ne i}^N(x_j-x_i)}\;,\label{heqsa} \\
\wt{h}_j^2&=&-p\frac{\prod_{i=1}^N(\wt{x}_j-x_i)}{\prod_{j\ne i}^N(\wt{x}_j-\wt{x}_i)} \;, \label{heqsb1} 
\end{eqnarray}
%
for $j=1,2,...,N$. Equating \eqref{heqsa} with \eqref{heqsb1}, we obtain the system of equations
\begin{equation}\label{eqmotion1}
-\frac{p }{\hypotilde 0 {p }}\frac{(x_i-\wt{x}_i)}{(x_i-{\hypotilde 0 x}_i)}=\prod\limits_{\mathop {j = 1}\limits_{j\ne i}}^N \frac{(x_i-{\hypotilde 0 x}_j)}{(x_i-\wt{x}_j)}\;.
\end{equation}
For simplicity, we take $p$ to be constant and then \eqref{eqmotion1} is simply the discrete-time equations of motion for the Calogero's goldfish system in the tilde-direction, see \cite{Suris}.  
%
\end{subequations}
\\
\\
\textbf{\emph{Second discrete flow}}:
We consider the compatibility between \eqref{LM12} and \eqref{LM13}
\begin{subequations}
\begin{eqnarray}\label{coN}
\wh{\boldsymbol L}_{\kappa}\boldsymbol N_{\kappa}&=&\boldsymbol N_{\kappa}\boldsymbol L_{\kappa} \nn\\
\left( \frac{\wh h\wh h^T}{\kappa}+\wh{\boldsymbol L}_0\right)\left( \frac{\wh{h}h^T}{\kappa}+\boldsymbol N_0\right)
&=&\left(\frac{\wh{h}h^T}{\kappa}+\boldsymbol N_0\right)\left( \frac{hh^T}{\kappa}+\boldsymbol L_0\right)\;.
\end{eqnarray}
What we obtain are the relation
 \begin{equation}\label{conh2}
 \sum_{j=1}^N \wh{h}_j^2=\sum_{j=1}^N h_j^2\;,
 \end{equation}
 and the set of equations
\begin{equation}\label{idenforh}
\sum_{j=1}^N
\frac{\wh{h}_j^2}{\wh{x}_j-{x}_l}
=\sum_{j=1}^N
\frac{h_j^2}{\wh{x}_i-x_j}\;,
\end{equation}
for all $i,j=1,2,...,N$. 
Using the same argument as in the previous case, we obtain
\begin{eqnarray}
\sum_{j=1}^N
\frac{\wh{h}_j^2}{\wh{x}_j-x_l}&=&-q\;,\;\;\;\;\forall l\;, \\
\sum_{j=1}^N
\frac{h_j^2}{\wh{x}_i-x_j}&=&-q\;,\;\;\;\;\forall i \;,
\end{eqnarray}
but with different parameter $q = q(m)$. Using the Lagrange interpolation formula, we get 
\begin{eqnarray}
h_j^2&=&q\frac{\prod_{i=1}^N(x_j-\wh{x}_i)}{\prod_{j\ne i}^N(x_j-x_i)}\;,\label{h21} \\
\wh{h}_j^2&=&-q\frac{\prod_{i=1}^N(\wh{x}_j-x_i)}{\prod_{j\ne i}^N(\wh{x}_j-\wh{x}_i)} \;, \label{h22}
\label{heqsb} 
\end{eqnarray}
for $i=1,2,...,N$, and a set of equations
\begin{equation}\label{eqmotion12}
-\frac{q }{\hypohat 0 {q}}\frac{(x_i-{\hypohat 0 x}_i)}{(x_i-\wh{x}_i)}=\prod\limits_{\mathop {j = 1}\limits_{j \ne i} }^N \frac{(x_i-\wh{x}_j)}{(x_i-{\hypohat 0 x}_j)}\;.
\end{equation}
\end{subequations}
We also take $q$ to be constant and then \eqref{eqmotion12} is again the discrete-time equations of motion for the Calogero's goldfish system in the hat-direction. 
%
\\
\\
\textbf{\emph{Commutativity between flows}}:
The last compatibility is between \eqref{LM13} and \eqref{LM14}.
\begin{subequations}
\begin{eqnarray}\label{coNM}
\wh{\boldsymbol M}_{\kappa}\boldsymbol N_{\kappa}&=&\wt{\boldsymbol N}_{\kappa}\boldsymbol M_{\kappa} \nn\\
\left( \frac{\wh{\wt h}\wh h^T}{\kappa}+\wh{\boldsymbol M}_0\right)\left( \frac{\wh{h}h^T}{\kappa}+\boldsymbol N_0\right)
&=&\left(\frac{\wh{\wt h}\wt h^T}{\kappa}+\wt{\boldsymbol N}_0\right)\left( \frac{\wt hh^T}{\kappa}+\boldsymbol M_0\right)\;.
\end{eqnarray}
Equation \eqref{coNM} gives the relation
\begin{equation}\label{conh3}
\sum_{j=1}^N \wt h_j^2=\sum_{j=1}^N\wh h_j^2\;,
\end{equation}
which can be considered as the consequence of the first two relations on the variable $h_i$. Furthermore, we have
\begin{eqnarray}
\wh{\wt{h}}\wh{h}^T\boldsymbol N_0-\wh{\wt{h}}\wt{h}^T\boldsymbol M_0&=&\wt{\boldsymbol N}_0\wt h h^T-\wh{\boldsymbol M}_0\wh h h^T\;,\\
\wh{\boldsymbol M}_0\boldsymbol N_0&=&\wt{\boldsymbol N}_0\boldsymbol M_0\;,
\end{eqnarray}
which produce a set of equations
\begin{equation}
\sum_{j=1}^N\left( \frac{\wh{h}_j^2}{\wh{\wt{x}}_i-\wh{x}_j}- \frac{\wt{h}_j^2}{\wh{\wt{x}}_i-\wt{x}_j}\right)=
\sum_{j=1}^N\left( \frac{\wt{h}_j^2}{{\wt{x}}_j-{x}_l}- \frac{\wh{h}_j^2}{\wh{{x}}_j-{x}_l}\right)\;.
\end{equation}
Again this equation is noting but the consequence of equations \eqref{heqsa} , \eqref{heqsb1}, \eqref{h21} and \eqref{h22}.
\\
\\
Equating \eqref{heqsa} with \eqref{h21} and \eqref{heqsb1} with \eqref{h22}, we obtain
%
\begin{eqnarray}
\frac{p}{q}&=&\prod_{j=1}^N\frac{(x_i-\wh{x}_j)}{(x_i-\wt{x}_j)}\;,\label{CON1a}\\
\frac{p}{q }&=&\prod_{j=1}^N\frac{(x_i-{\hypohat 0 x}_j)}{(x_i-{\hypotilde 0 x}_j)}\;.\label{CON1b}
\end{eqnarray}
%
Using equations of motion \eqref{eqmotion1} and \eqref{eqmotion12}, we have another two relations
\begin{eqnarray}
-\frac{p}{q}&=&\prod_{j=1}^N\frac{(x_i-\wh{x}_j)}{(x_i-{\hypotilde 0 x}_j)}\;,\label{CON2a}\\
-\frac{p}{q }&=&\prod_{j=1}^N\frac{(x_i-{\hypohat 0 x}_j)}{(x_i-\wt{x}_j)}\;.\label{CON2b}
\end{eqnarray}
These equations can be treated as constraints describing how two discrete flows connect at the centre of the lattice as shown in figure \ref{lattice}.
\begin{figure}[h!]
   \centering 
    \includegraphics[width=0.50\textwidth]{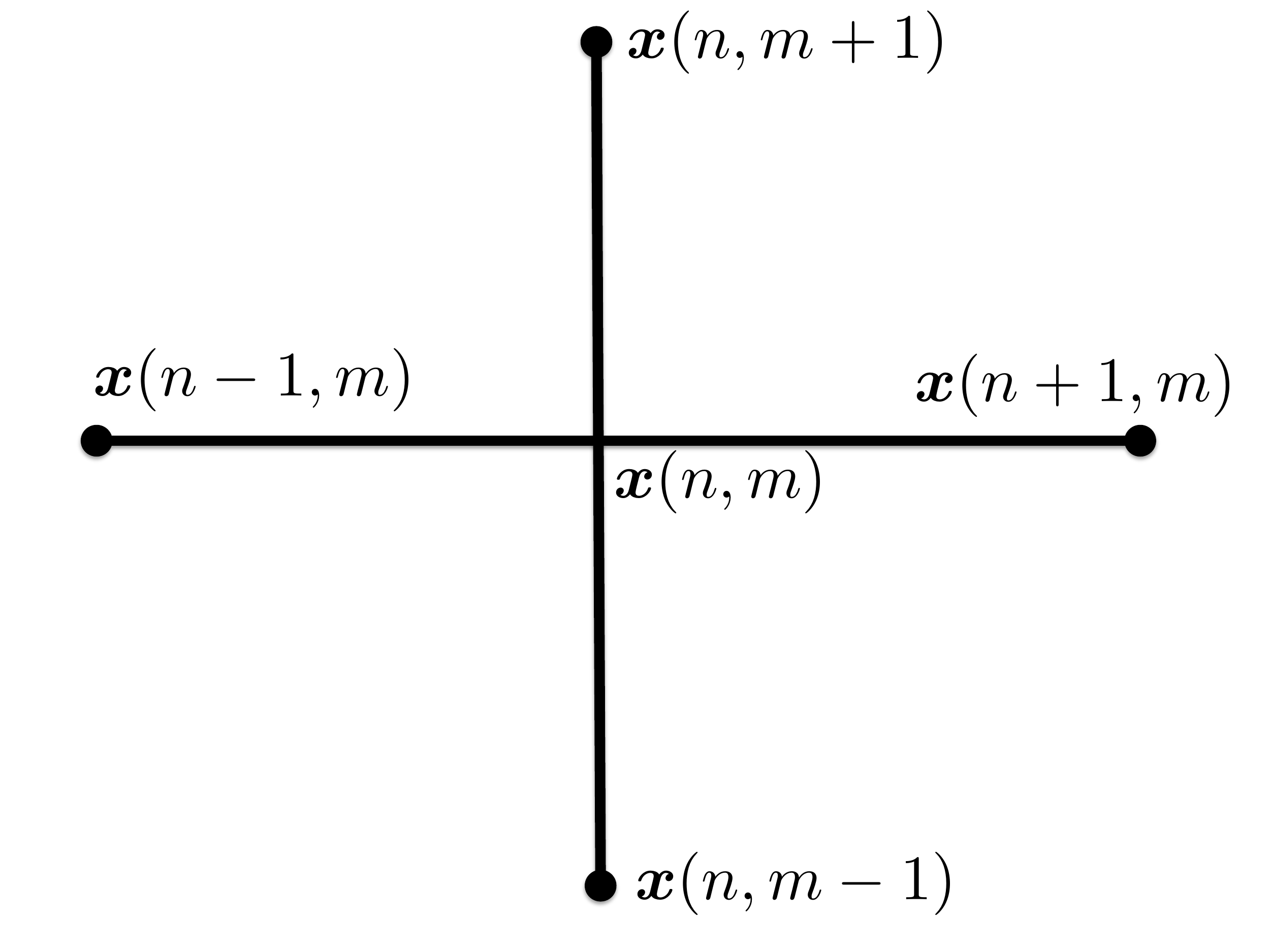} 
   \caption{\small{The lattice structure is constituted from two discrete flows: $(n,m)$. The horizontal and vertical lines are described by the equations of motion for tilde-direction and hat-direction, respectively. The constraints describe the relation between two discrete flows at four corners around the centre. }} 
   \label{lattice}
\end{figure}
%
\end{subequations}
Equating between \eqref{CON1a} and \eqref{CON1b} as well as \eqref{CON2a} and \eqref{CON2b} give equation of motion for discrete-time Calogero's goldfish
\begin{equation}\label{eq12}
\prod_{j=1}^N\frac{(x_i-\wt{x}_j)}{(x_i-{\hypotilde 0 x}_j)}=
\prod_{j=1}^N\frac{(x_i-\wh{x}_j)}{(x_i-{\hypohat 0 x}_j)}\;,
\end{equation}
which expresses the compatibility with the set of O$\Delta$Es.
\\
\\
\textbf{\emph{Exact solution}}:
We first start to consider the solution for the tilde-direction. The matrices $\boldsymbol{M}_0$ and $\boldsymbol{L}_0$ can be rewritten in the form
\begin{subequations}
\begin{eqnarray}
\wt{\boldsymbol X}\boldsymbol{M}_0-\boldsymbol{M}_0\boldsymbol X=\wt{h}h^T\;,\label{MX}\\
\boldsymbol{L}_0=hh^T\;,\label{LX}
\end{eqnarray}
\end{subequations}
where $\boldsymbol X=\sum_{i=1}^Nx_iE_{ii}$ is the diagonal matrix. 
From the Lax equation \eqref{iden1} and \eqref{iden2}, we obtain the relations
\begin{subequations}\label{XX1}
\begin{eqnarray}
\wt{\boldsymbol{L}}_0\boldsymbol{M}_0&=&\boldsymbol{M}_0\boldsymbol{L}_0\;,\label{LM1}\\
\wt{\boldsymbol{L}}_0\wt{h}-\boldsymbol{M}_0h&=&-p\wt{h}\;,\label{LM2}\\
h^T\boldsymbol{L}_0-\wt{h}^T\boldsymbol{M}_0&=&-ph^T\;.\label{LM3}
\end{eqnarray}
\end{subequations}
We now factorise the Lax matrices as follows: 
\begin{equation}\label{XX2}
\boldsymbol{L}_0=\boldsymbol U\bLam \boldsymbol U^{-1}\;,\;\;\;\;\;\mbox{and}\;\;\;\;\;\boldsymbol{M}_0=\wt{\boldsymbol U}\boldsymbol U^{-1}\;,
\end{equation}
where $\boldsymbol U$ is an invertible $N \times N$ matrix and the matrix $\bLam$ is constant: $\wt{\bLam}=\bLam$. Obviously, if $\boldsymbol{L}_0$ 
is diagonalisable $\bLam$ is just its diagonal matrix of eigenvalues. Next, let us introduce
\begin{equation}\label{Yrs}
\boldsymbol{Y}=\boldsymbol U^{-1}\boldsymbol{X}\boldsymbol U\;,\;\;\;\boldsymbol{r}=\boldsymbol U^{-1}\cdot h\;,\;\;\;\boldsymbol{s}^T=h^T\cdot \boldsymbol U\;,
\end{equation}
 and we get from \eqref{XX1} and \eqref{XX2},
\begin{equation}\label{rs}
(pI+\bLam)\cdot \wt{\boldsymbol r}=\boldsymbol r\;,\;\;\;\;\boldsymbol{s}^T\cdot (\mathrm{p}I+\bLam)=\wt{\boldsymbol s}^T\;,
\end{equation}
where $I$ is the unit matrix. From \eqref{MX}
\begin{eqnarray}\label{Yrs2}
\wt{\boldsymbol Y}-\boldsymbol Y&=&\wt{\boldsymbol r}\boldsymbol s^T\;.
\end{eqnarray}
The dyadic $\boldsymbol r\boldsymbol s^T$ can be eliminated from \eqref{Yrs2} by making use of \eqref{rs} resulting to
\begin{equation}\label{Yn1}
\wt{\boldsymbol{Y}}= \boldsymbol Y +\frac{\bLam}{(p\boldsymbol{I}+\bLam)}\;.
\end{equation}
After $n$ discrete steps, we find that
\begin{equation}\label{Yn22}
\boldsymbol{Y}(n,m)= \boldsymbol Y(0,m)+\frac{n\bLam}{(p\boldsymbol{I}+\bLam)}\;,
\end{equation}
Automatically, we find that the solution in the hat-direction is
\begin{eqnarray}\label{Ynm1}
\boldsymbol{Y}(n,m)= \boldsymbol Y(n,0)+\frac{m\bLam}{(q\boldsymbol{I}+\bLam)}\;.
\end{eqnarray}
Combining \eqref{Yn22} and \eqref{Ynm1}, we obtain the complete solution of the system
\begin{eqnarray}\label{EXACT}
\boldsymbol{Y}(n,m)= \boldsymbol Y(0,0)+\frac{n\bLam}{(p\boldsymbol{I}+\bLam)}+\frac{m\bLam}{(q\boldsymbol{I}+\bLam)}\;,
\end{eqnarray} 
where $x_i(n,m)$ can be determined by considering the eigenvalues of the matrix $\boldsymbol{Y}(n,m)$.
\\\\
\textbf{\emph{Discrete actions}}:
We find that the equations of motion in the tilde-direction (horizontal discrete curve in figure \ref{lattice}) are the consequence of variation of the discrete action
 \begin{subequations}
\begin{eqnarray}
{S}_{H}&=&\mathscr{L}_{(p)}(\boldsymbol{x},\wt{\boldsymbol{x}})+\mathscr{L}_{(p)}(\boldsymbol{x},{\hypotilde 0 {\boldsymbol{x}}})\;,\label{actionn}
\end{eqnarray}
yielding
\begin{eqnarray}\label{LT}
\delta S_{H}=0\;\;\Rightarrow\;\; \frac{\partial\mathscr{L}_{(p)}}{\partial{\wt{x}_i}}+\wt{\frac{\partial\mathscr{L}_{(p)}}{\partial{x_i}}}=0\;,
\end{eqnarray}
where
\begin{eqnarray}
\mathscr{L}_{(p)}&=&\sum_{i,j=1}^N(x_i-\wt{x}_j)\ln(x_i-\wt{x}_j)+\ln\left|p\right|\sum_{i=1}^N(x_i -\wt{x }_i) \;.\label{2Lagn}
\end{eqnarray}
Equation \eqref{LT} gives the discrete-time equations of motion in the tilde-direction equation \eqref{eqmotion1}.
In the hat-direction (vertical discrete curve in figure \ref{lattice}), we also have the equations of motion which are the consequence of variation of the discrete action
\begin{eqnarray}
{S}_{V}&=&\mathscr{L}_{(q)}(\boldsymbol{x},\wh{\boldsymbol{x}})+\mathscr{L}_{(q)}(\boldsymbol{x},{\hypohat 0 {\boldsymbol{x}}})\;,\label{actionn}
\end{eqnarray}
yeilding
\begin{eqnarray}\label{LH}
\delta S_{V}=0\;\;\Rightarrow\;\; \frac{\partial\mathscr{L}_{(q)}}{\partial{\wh{x}_i}}+\wh{\frac{\partial\mathscr{L}_{(q)}}{\partial{x_i}}}=0\;,
\end{eqnarray}
where
\begin{eqnarray}
\mathscr{L}_{(q)}&=&\sum_{i,j=1}^N\left(x_i-\wh{x}_j\right)\ln(x_i-\wh{x}_j)+\ln\left|q \right|\sum_{i=1}^N(x_i -\wh{x}_i)\;.\label{2Lagm}
\end{eqnarray}
Equation \eqref{LH} gives the discrete-time equations of motion in the tilde-direction equation \eqref{eqmotion12}.
Furthermore, we also have another four discrete actions corresponding to two different discrete curves connecting at the centre as shown in figure \ref{lattice2}(a)
\begin{figure}[h!]
  \subfigure[]{
   \includegraphics[scale =.3] {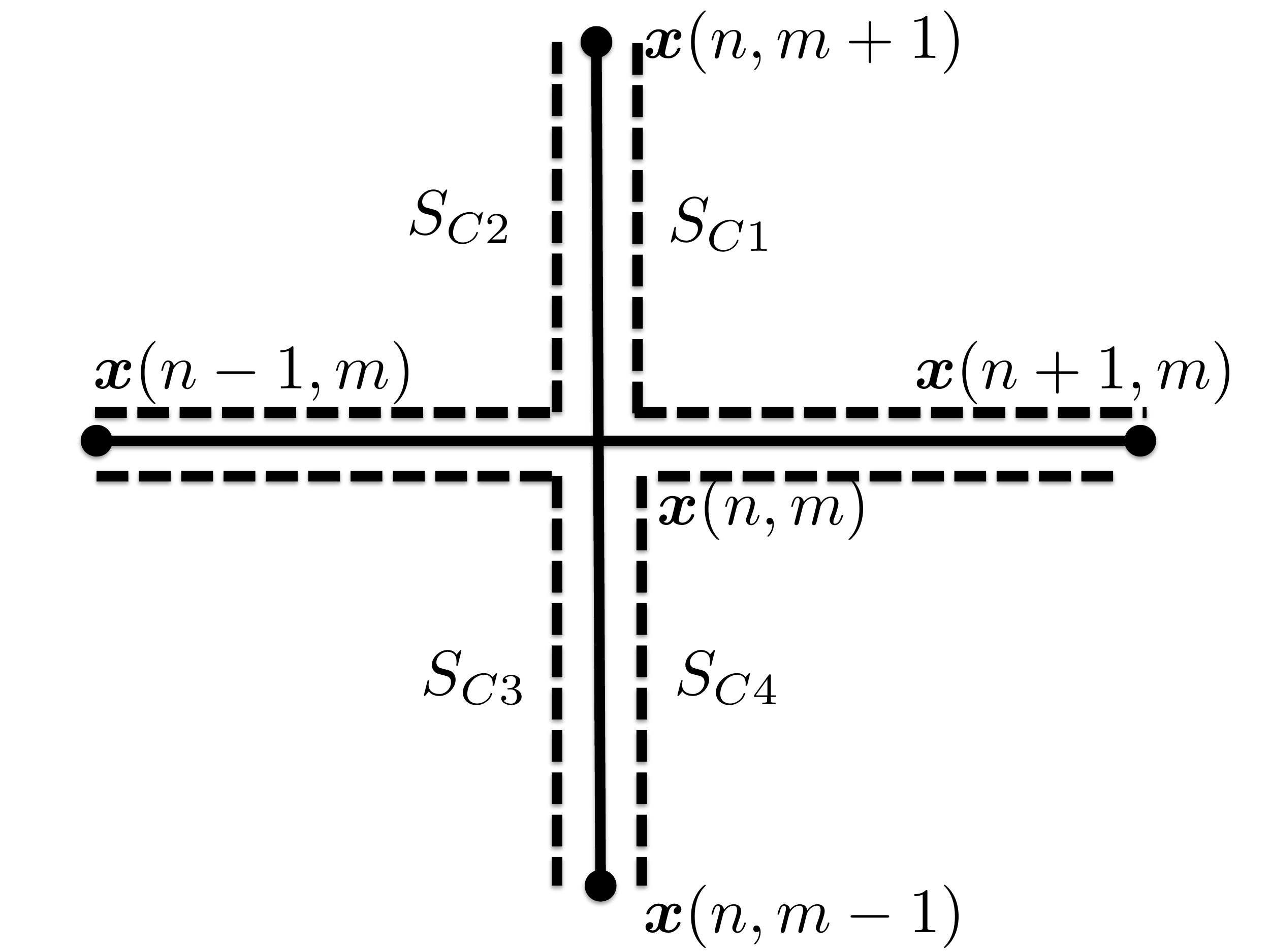}
   \label{fig:subfig1}
 }
 \subfigure[]{
   \includegraphics[scale =.3] {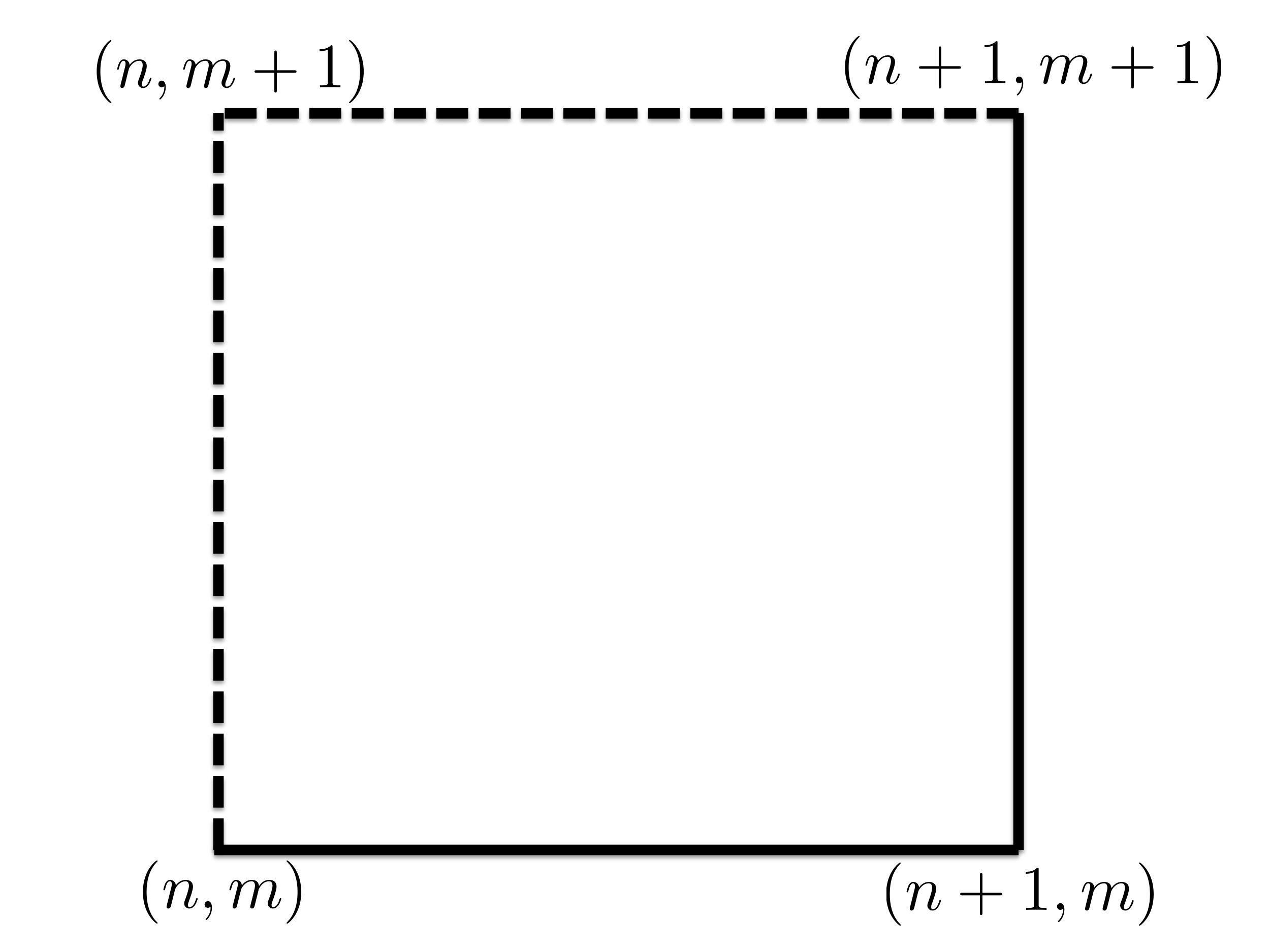}
   \label{fig:subfig2}
 }

   \caption{\small{(a) Discrete actions around the centre of the lattice. (b) The deformation of the discrete curve on the space of the independent variables. }} 
   \label{lattice2}
\end{figure}
\begin{eqnarray}
S_{C1}&=&\mathscr{L}_{(p)}(\wt{\boldsymbol{x}},\boldsymbol{x})+\mathscr{L}_{(q)}(\boldsymbol{x},\wh{\boldsymbol{x}})\;,\\
S_{C2}&=&\mathscr{L}_{(p)}({\hypotilde 0 {\boldsymbol{x}}},\boldsymbol{x})+\mathscr{L}_{(q)}(\boldsymbol{x},\wh{\boldsymbol{x}})\;,\\
S_{C3}&=&\mathscr{L}_{(p)}({\hypotilde 0 {\boldsymbol{x}}},\boldsymbol{x})+\mathscr{L}_{(q)}(\boldsymbol{x},{\hypohat 0 {\boldsymbol{x}}})\;,\\
S_{C4}&=&\mathscr{L}_{(p)}(\wt{\boldsymbol{x}},\boldsymbol{x})+\mathscr{L}_{(q)}(\boldsymbol{x},{\hypohat 0 {\boldsymbol{x}}})\;.
\end{eqnarray}
The variation on these four actions yields nothing but the constraint equations.
\\
\\
Another important feature for this discrete Lagrangians is the closure relation 
\begin{equation}\label{closure}
 \widehat{\mathscr{L}_{(p)}(\boldsymbol{x},\widetilde{\boldsymbol{x}})} - \mathscr{L}_{(p)}(\boldsymbol{x},\widetilde{\boldsymbol{x}}) -
 \widetilde{\mathscr{L}_{(q)}(\boldsymbol{x},\widehat{\boldsymbol{x}})} + \mathscr{L}_{(q)}(\boldsymbol{x},\widehat{\boldsymbol{x}}) = 0\;,
\end{equation}
which is the direct result of variation of the discrete curve on the space of independent variables $(n,m)$, see also \cite{Sikarin1, Sikarin2}. The validity of \eqref{closure} can be shown with the help of equations of motion \eqref{eqmotion1} and \eqref{eqmotion12}. The closure relation ensures that the action of the system is invariant under the local deformation of the discrete curve, see figure \ref{lattice2}(b).
\end{subequations}
\\
\\
\textbf{\emph{Remark}}: From the Lagrangians \eqref{2Lagn} and \eqref{2Lagm}, we define the momentum variables
\begin{eqnarray}
\texttt{p}_i=-\frac{\partial \mathscr{L}_{(p)}}{\partial \wt x_i}=\sum_{j=1}^N\ln(x_j-\wt x_i)+1+\ln|p|\;,\label{p1}\\
\pi_i=-\frac{\partial \mathscr{L}_{(q)}}{\partial \wh x_i}=\sum_{j=1}^N\ln(x_j-\wh x_i)+1+\ln|q|\;,\label{pi}
\end{eqnarray}
corresponding to the tilde-direction and the hat-direction, respectively. Using the above relations, we can write \eqref{heqsa} in the form
\begin{eqnarray}
h_k^2=\frac{e^{\texttt{p}_k-1}}{\prod_{j \ne k}(x_k-x_j)}\;,
\end{eqnarray}
and \eqref{h21} in the form
\begin{eqnarray}
h_k^2=\frac{e^{\pi_k-1}}{\prod_{j \ne k}(x_k-x_j)}\;.
\end{eqnarray}
The Hamiltonian of the system is given by
\begin{equation}
\mathscr{H}=\mbox{Tr}\boldsymbol{L}_0=\sum_{k=1}^N h_k^2\;.
\end{equation}
Then equations \eqref{conh}, \eqref{conh2} and \eqref{conh3} are
\begin{eqnarray}
&&\sum_{k=1}^Nh_k^2=\sum_{k=1}^N\wt h_k^2\;\;\;\mapsto\;\;\;\mathscr{H}=\wt{\mathscr{H}}\;,\label{h1}\\
&&\sum_{k=1}^Nh_k^2=\sum_{k=1}^N\wh h_k^2\;\;\;\mapsto\;\;\;\mathscr{H}=\wh{\mathscr{H}}\;,\label{h2}\\
&&\sum_{k=1}^N\wt h_k^2=\sum_{k=1}^N\wh h_k^2\;\;\;\mapsto\;\;\;\wt{\mathscr{H}}=\wh{\mathscr{H}}\;.\label{h3}
\end{eqnarray}
Equations \eqref{h1} and \eqref{h2} represent the energy conservation law under the tilde-direction and the hat-direction since both discrete time flows share the same $\boldsymbol{L}$ matrix. Equation \eqref{h3} can be treated as the discrete analogue of the commuting flows.

\section{The partial-continuum limit}\label{skewlimit}
\setcounter{equation}{0} 
In this section, we consider the continuum limit of the discrete-time Calogero's goldfish system which had been investigated in the previous section. Since there are two discrete-time variables $(n,m)$, we may perform directly continuum limit of each of these variables resulting the usual continuous-time Calogero's goldfish system \cite{Suris}. We now work with another type of continuum limit namely the \emph{skew limit}. In order to proceed to this limit, we introduce a new discrete-time variable $N=n+m$ and with this new variable we have a set of transformations on the variables such that
\begin{eqnarray}\label{changingvariables}
x(n,m)&\mapsto& \mathsf x(\mathsf N,m)=:{\mathsf x}\nn\\
\widetilde x = x(n+1,m) &\mapsto & \mathsf x(\mathsf N+1,m)=:\bar{\mathsf x}\;, \nn\\
\widehat x = x(n,m+1) &\mapsto & \mathsf x(\mathsf N+1,m+1)=:\widehat{\bar{ {\mathsf x}}}\;,\nn\\ 
{\widehat{\widetilde x}} = x(n+1,m+1) &\mapsto & \mathsf x(\mathsf N+2,m+1)=:\widehat{\bar{\bar {\mathsf x}}}\;.\nn
\end{eqnarray}
We also introduce $\varepsilon=p-q$ and $\varepsilon m=\tau$ and then send $n\rightarrow -\infty$, $m \rightarrow \infty$, $\varepsilon \rightarrow 0$ while keeping $\mathsf N$ and $\tau$ fixed.
\\
\\
We first consider the skew limit on the exact solution given in \eqref{EXACT}. We can rewrite the solution in terms of the new variables
\begin{eqnarray}\label{EXACTSKEW}
\mathsf {\boldsymbol Y}(n,m)\mapsto \mathsf {\boldsymbol Y}(\mathsf {N},m)&=&\boldsymbol Y(0,0)+\frac{\mathsf{N}\bLam}{pI+\bLam}+\frac{m\varepsilon\bLam}{(p+\bLam)^2(1-\frac{\varepsilon}{p+\bLam})}\;.\\
\lim\limits_{\mathop {m \rightarrow \infty} \limits_{\mathop {\varepsilon \rightarrow 0}\limits_{\varepsilon
m \rightarrow \tau}} }\mathsf {\boldsymbol Y}(\mathsf {N},m)\mapsto \mathsf {\boldsymbol Y}(\mathsf {N},\tau)&=&\boldsymbol Y(0,0)+\frac{\mathsf{N}\bLam}{pI+\bLam}+
\frac{\tau\bLam}{(pI+\bLam )^2}\;.\label{Y1}
\end{eqnarray} 
The shift on the position of particles in the hat-direction becomes
\begin{eqnarray}
\widehat{\bar{ {\mathsf x}}}=\mathsf x(\mathsf N+1,m+1)&\mapsto &\wh{ \bar{ \mathrm{x}}}=\mathrm{x}(\mathsf N+1,\tau +\varepsilon )\nn \\
\hypohat 0 {\underline{ \mathrm{x}}}=\mathsf x(\mathsf N-1,m-1)&\mapsto &\hypohat 0 {\underline{ \mathrm{x}}}=\mathrm{x}(\mathsf N-1,\tau -\varepsilon)\nn
\end{eqnarray}
and the expansions with respect to $\varepsilon$ lead to
\begin{eqnarray}\label{Taylor}
\wh{ \bar{ \mathrm{x}}}=\mathrm{x}(\mathsf N+1,\tau +\varepsilon )&\mapsto&\bar{ \mathrm{x}}+ \varepsilon\frac{\partial \bar{ \mathrm{x}}}{\partial {\tau}}+\frac{\varepsilon^2}{2}\frac{\partial^2 {\bar{ \mathrm{x}}}}{\partial {\tau^2}}+ ...\;,\\
\hypohat 0 {\underline{ \mathrm{x}}}=\mathrm{x}(\mathsf N-1,\tau -\varepsilon)&\mapsto&\underline{ \mathrm{x}}- \varepsilon\frac{\partial \underline{ \mathrm{x}}}{\partial {\tau}}+\frac{\varepsilon^2}{2}\frac{\partial^2 {\underline{ \mathrm{x}}}}{\partial {\tau^2}}- ...\;.
\end{eqnarray}
The positions of the particles $ \mathrm{x}(\mathsf N,\tau)$ can be computed by considering the eigenvalues of \eqref{Y1} \cite{Niff}.
%
%
%
%
%
\\\\
\textbf{\emph{Equations of motion and constraints}}:
The equations of motion \eqref{eqmotion12} in terms of new variables $(\mathsf N,\tau )$ are given by
\begin{subequations}
\begin{eqnarray}
\sum_{j=1}^N\left(\ln(\wh{\bar{\mathrm  x}}_j-\mathrm  x_i)-\ln(\mathrm x_i-\underline {\hypohat 0 {\mathrm {x}}}_j )\right) &=& 0 \;.\label{eqln2skew}
\end{eqnarray}
Expanding the variable $\mathrm x$ with respect to the variable $\varepsilon$ and collecting terms in power of $\varepsilon$, we find
%
%
%
%
\begin{eqnarray}\label{eq12}
&&\mathcal O(\varepsilon^0 ):\;\;\;\;\;\sum_{j=1}^N\left(\ln({\bar{\mathrm  x}}_j-\mathrm  x_i)-\ln(\mathrm x_i-\underline  {\mathrm {x}}_j )\right) = 0\;,\label{x11}\\
&&\mathcal O(\varepsilon^1 ):\;\;\;\;\;\sum_{j=1}^N\left[ \frac{\partial {\bar {\mathrm x}_j}}{\partial \tau}\left( \frac{1}{\bar{\mathrm x}_j-\mathrm x_i}\right)-\frac{\partial \underline {\mathrm x}_j}{\partial \tau}\left( \frac{1}{\mathrm x_i-\underline {\mathrm x}_j}\right)\right]=0\;.\label{x12}
\end{eqnarray}
We terminate the series at $\mathcal O(\varepsilon^1)$, but the higher order terms can be directly obtained by continuing the expansion. What we see from the result is that \eqref{x11} the equations of motion of Calogero's goldfish system in terms of the new discrete-time variable $\mathsf N$. The \eqref{x12} is the equations of motion of Calogero's goldfish system in terms of the continuous variable $\tau$. 
\\
\\
Next, we perform the limit on the constraints and collect the first dominant terms
%
\begin{eqnarray}
-\frac{1}{p} &=&\sum_{j=1}^N\frac{\partial {\bar{\mathrm x}_j}}{\partial \tau}\left(\frac{1}{\mathrm x_i-\bar{\mathrm x}_j} \right)\;,\label{CONSKEW1}\\
\frac{1}{p} &=&\sum_{j=1}^N\frac{\partial {\underline {\mathrm x}_j}}{\partial \tau}\left(\frac{1}{\mathrm x_i-\underline {\mathrm x}_j} \right)\;.\label{CONSKEW2}
\end{eqnarray}
\end{subequations}
The combination of \eqref{CONSKEW1} with \eqref{CONSKEW2} gives directly the equations of motion \eqref{x12}.
\\\\
\textbf{\emph{The Lagrangians and closure relation}}:
We start in this section to write Lagrangian \eqref{2Lagm} in terms of the variables $(\mathsf{N},m)$
\begin{subequations}
\begin{eqnarray}
\mathscr{L}_{(q)}&\mapsto&\sum_{i,j=1}^N\left(\mathsf x_i-\wh{\bar{\mathsf x}}_j\right)\ln(\mathsf x_i-\wh{\bar{\mathsf x}}_j)+\ln\left|p-\varepsilon \right|\sum_{i=1}^N(\mathsf x_i -\wh{\bar{\mathsf x}}_i)\;.\label{2Lagm2}
\end{eqnarray}
Then, we expand with respect to the variable $\varepsilon$ resulting to
\begin{eqnarray}
\mathscr{L}_{(q)}\mapsto  \varepsilon^0 \mathcal{L}_{(\mathsf N)}+ \varepsilon^1 \mathcal{L}_{(\tau)}^{(1)}  
+...\;,\label{2Lagmv}
\end{eqnarray}
where 
\begin{eqnarray}
\mathcal{L}_{(\mathsf N)}&=&\sum_{i,j=1}^N\left(\mathsf x_i-\bar{\mathsf x}_j\right)\ln(\mathsf x_i-\bar{\mathsf x}_j)+\ln\left|p \right|\sum_{i=1}^N(\mathsf x_i -\bar{\mathsf x}_i)\;,\label{2Lagm22}\\
\mathcal{L}_{(\tau)}^{(1)}&=& -\sum_{i,j=1}^N\frac{\partial {\bar {\mathsf x}_j}}{\partial \tau}(1+\ln(\mathsf x_i-\bar{\mathsf x}_j))-\ln \left|p \right| \sum_{i=1}^N\frac{\partial\bar{\mathsf x }_i}{\partial \tau}-\frac{1}{p}\sum_{i=1}^N(\mathsf x_i -\bar{\mathsf x}_i)\label{2Lagma} \;.
\end{eqnarray}
These Lagrangians gives the equations of motion \eqref{x11} and \eqref{x12}. This can be seen by substituting the Lagrangians in the following Euler-Lagrange equations
\begin{eqnarray}
\frac{\partial\mathcal{L}_{(\mathsf N)}}{\partial \mathsf x}+\underline{\frac{\partial\mathcal{L}_{(\mathsf N)}}{\partial \bar{\mathsf x}}}&=&0\;,\\
\frac{\partial\mathcal{L}_{(\tau)}^{(1)}}{\partial \mathsf x}
+\underline{\frac{\partial\mathcal{L}_{(\tau)}^{(1)}}{\partial \bar{\mathsf x}}}
-\underline{\frac{d}{d\tau}\left(\frac{\partial\mathcal{L}_{(\tau)}^{(1)}}{\partial \left(\frac{\partial\bar{\mathsf x}}{\partial\tau}\right)}\right)}&=&0\;.
\end{eqnarray}
Furthermore, the constraints \eqref{CONSKEW1} and \eqref{CONSKEW2} are the result of Euler-Lagrange of Lagrangian $\mathcal{L}_{(\tau)}^{(1)}$ with respect to the variable $\bar{\mathsf x}$
\begin{eqnarray}
\frac{\partial\mathcal{L}_{(\tau)}^{(1)}}{\partial \bar{\mathsf x}}-\frac{d}{d\tau}\left(\frac{\partial\mathcal{L}_{(\tau)}^{(1)}}{\partial \left(\frac{\partial\bar{\mathsf x}}{\partial\tau}\right)}\right)&=&0\;.
\end{eqnarray}
\\
Next, we perform the continuum limit on the closure relation and collect for the first two dominant terms in power of $\varepsilon$
\begin{eqnarray}
&&\mathcal O(\varepsilon^1 ):\;\;\;\;\;\frac{\partial \mathcal{L}_{(\mathsf N)}}{\partial \tau}=\mathcal{L}_{(\tau)}^{(1)}-\underline {\mathcal{L}_{(\tau)}^{(1)}}\;.\label{CL1}
\end{eqnarray}
The equation \eqref{CL1} represents the closure relation between the discrete Lagrangian $\mathcal{L}_{(\mathsf N)}$ and continuous Lagrangian $\mathcal{L}_{(\tau)}^{(1)}$. This relation guarantees the invariance of the action on the space of independent variables mixing between discrete variable $\mathsf N$ and continuous variable $\tau$.

\end{subequations}


\section{The full continuum limit}\label{fullLIMIT}
In this section, we perform the remaining task in order to complete the continuum limit. We set out with the expansion of
\eqref{Y1} with respect to the variable $p$
\begin{eqnarray}\label{fulllimitsolution}
\boldsymbol{\mathsf {Y}}(\mathsf {N},\tau)&\mapsto
& \boldsymbol{\mathsf Y}(0,0)+\mathsf{ N}\frac{\boldsymbol{\Lambda}}{p}\left(1-\frac{\boldsymbol{\Lambda}}{p}+\left(\frac{\boldsymbol{\Lambda}}{p}\right)^{2}-\left(\frac{\boldsymbol{\Lambda}}{p}\right)^{3}+...\right)\;\nn\\
&& +\tau\boldsymbol{\Lambda}\left(\frac{1}{p^2}-\frac{2\boldsymbol{\Lambda}}{p^3}+\frac{3\boldsymbol{\Lambda}^2}{p^4}-...\right)\;,
\end{eqnarray}
and then we collect terms in power of $\boldsymbol{\Lambda}$
\begin{eqnarray}\label{Yt}
\boldsymbol{\mathsf {Y}}(\mathsf {N},\tau)&\mapsto
&\boldsymbol{\mathsf {Y}}(t_1,t_2,t_3,...,t_N)=\boldsymbol{\mathsf Y}(0,0)+\boldsymbol{\Lambda}t_1+\boldsymbol{\Lambda}^2t_2+\boldsymbol{\Lambda}^3t_3+...+\boldsymbol{\Lambda}^Nt_N\;,\nn\\
\end{eqnarray}
where
\begin{eqnarray}\label{t1t2t3}
t_1=\frac{\tau}{p^2}+\frac{\mathsf {N}}{p}\;,\;
 t_2=-\frac{2\tau}{p^3}-\frac{\mathsf {N}}{p^2}\;,\;
.....\;,\;t_N=(-1)^{N+1}\left(\frac{N\tau}{p^{N+1}}+\frac{\mathsf {N}}{p^N}\right)\;.
\end{eqnarray}
The position of the $i$th particle $X_i(t_1,t_2,...,t_N)$ can be determined by looking for the eigenvalues of \eqref{Yt}.
\\
\\
With these new continuous variables, we find that
\begin{eqnarray}\label{eq:5.4}
\frac{\partial{\mathsf x}_i}{\partial \tau}&=&\frac{\partial X_i}{\partial t_1}
\frac{\partial t_1}{\partial \tau}+\frac{\partial X_i}{\partial t_2}\frac{\partial t_2}{\partial \tau}+\frac{\partial X_i}{\partial t_3}
\frac{\partial t_3}{\partial \tau}+...+\frac{\partial X_i}{\partial t_N}
\frac{\partial t_N}{\partial \tau}\nonumber\\
&=&\frac{1}{p^2}\frac{\partial X_i}{\partial t_1}-\frac{2}{p^3}\frac{\partial X_i}{\partial t_2}+\frac{3}{p^4}\frac{\partial X_i}{\partial t_3}+...+\frac{(-1)^{N+1}N}{p^{N+1}}\frac{\partial X_i}{\partial t_N}\;,
\end{eqnarray}
and
\begin{eqnarray}\label{eq:5.5}
 {\mathsf x}_i(\mathsf {N}\pm 1)&=&e^{\pm\frac{\partial}{p\partial t_1}\mp\frac{\partial}{p^2\partial t_2}\pm\frac{\partial}{p^3\partial t_3}\mp\frac{\partial}{p^4\partial t_5}\pm....}X_i\;\nn\\
 &=&X_i \pm \frac{1}{p}\frac{\partial X_i}{\partial t_1} 
+\frac{1}{p^2}\left(\frac{1}{2}\frac{\partial^2 X_i}{\partial t_1^2} \mp \frac{\partial X_i}{\partial t_2}\right)
+\frac{1}{p^3}\left(\pm \frac{1}{6}\frac{\partial^3 X_i}{\partial t_1^3}-\frac{\partial^2 X_i}{\partial t_1\partial t_2} 
\pm \frac{\partial X_i}{\partial t_4}\right)\nn\\
&&+\frac{1}{p^4}\left(\frac{\partial X_i}{\partial t_3}\mp \frac{1}{2}\frac{\partial^3 X_i}{\partial t_1^2\partial t_2}+\frac{1}{2}\frac{\partial^2 X_i}{\partial t_2^2}+\frac{\partial^2 X_i}{\partial t_1\partial t_3}\right)
+\mathcal O(1/p^5)\;.
\end{eqnarray}
Later in this section, we restrict to the case of the first two time variables for the sake of simplicity.
\\\\
\textbf{\emph{Equations of motion}}:
Performing the expansion in \eqref{x11}, we find
\begin{eqnarray}\label{EQMOTIONRS1}
&&\mathcal O(1/p):\;\;\;\;
\frac{\partial^2 X_i}{\partial t_1^2}- 2\sum_{j=1,j\neq i}^N\frac{\partial X_i}{\partial t_1} \frac{\partial X_j}{\partial t_1}\frac {1}{X_i-X_j}= 0\;,\label{EqG1}\\
&&\mathcal O(1/p^2):\;\;\;\;\frac{\partial^2 X_i}{\partial t_1\partial t_2}
-2\sum_{j=1,j\neq i}^N\frac{\partial X_i}{\partial t_1}\frac{\partial X_j}{\partial t_2} \frac{1}{X_i-X_j}=0\;.\label{EqG2}
\end{eqnarray}
Eq. \eqref{EqG1} is just the usual equations of motion for the Calogero's goldfish system. Eq. \eqref{EqG2} can be considered to be the equations of motion of the system next in the hierarchy. The rest of equations of motion in the hierarchy can be determined by just pushing further on the expansion.
\\\\
\textbf{\emph{ Lagrangians}}:
We immediately observe that the Lagrangians corresponding the equations of motion \eqref{EqG1} and \eqref{EqG2} are
\begin{eqnarray}\label{eq:5.10}
L_{(t_1)}&=&\sum\limits_{i=1}^N\frac{\partial X_i}{\partial t_1}\ln\left| \frac{\partial X_i}{\partial t_1}\right|+\sum\limits_{i \ne j}^N \frac{\partial X_j}{\partial t_1}\ln|X_i-X_j|\;, \label{La1}\\
L_{(t_2)}&=&\sum\limits_{i=1}^N\left(\frac{\partial X_i}{\partial t_2}\ln\left| \frac{\partial X_i}{\partial t_1}\right|-\frac{1}{2}\frac{\partial X_i}{\partial t_2}\right)+\sum\limits_{i \ne j}^N\frac{\partial X_j}{\partial t_2} \ln|X_i-X_j|
\;,\label{La2}
\end{eqnarray}
with the Euler-Lagrnge equations
\begin{eqnarray}\label{eq:5.10c}
&&\frac{\partial  L_{(t_1)}}{\partial X_i}-\frac{\partial}{\partial t_1}\left( \frac{\partial 
 L_{(t_1)}}{\partial(\frac{\partial X_i}{\partial t_1})}\right)=0\;, \\
 &&\frac{\partial  L_{(t_2)}}{\partial X_i}-\frac{\partial}{\partial t_2}\left( \frac{\partial 
 L_{(t_2)}}{\partial(\frac{\partial X_i}{\partial t_2})}\right)=0\;.
\end{eqnarray}
Alternatively, Lagrangians \eqref{La1} and \eqref{La2} can be obtained by performing the full continuum limit on the action as in the case of Calogero-Moser system and Ruijsenaars-Schneider system, see  \cite{Sikarin1, Sikarin2}.
%
%
%
\begin{figure}[h!]
   \centering 
    \includegraphics[width=0.50\textwidth]{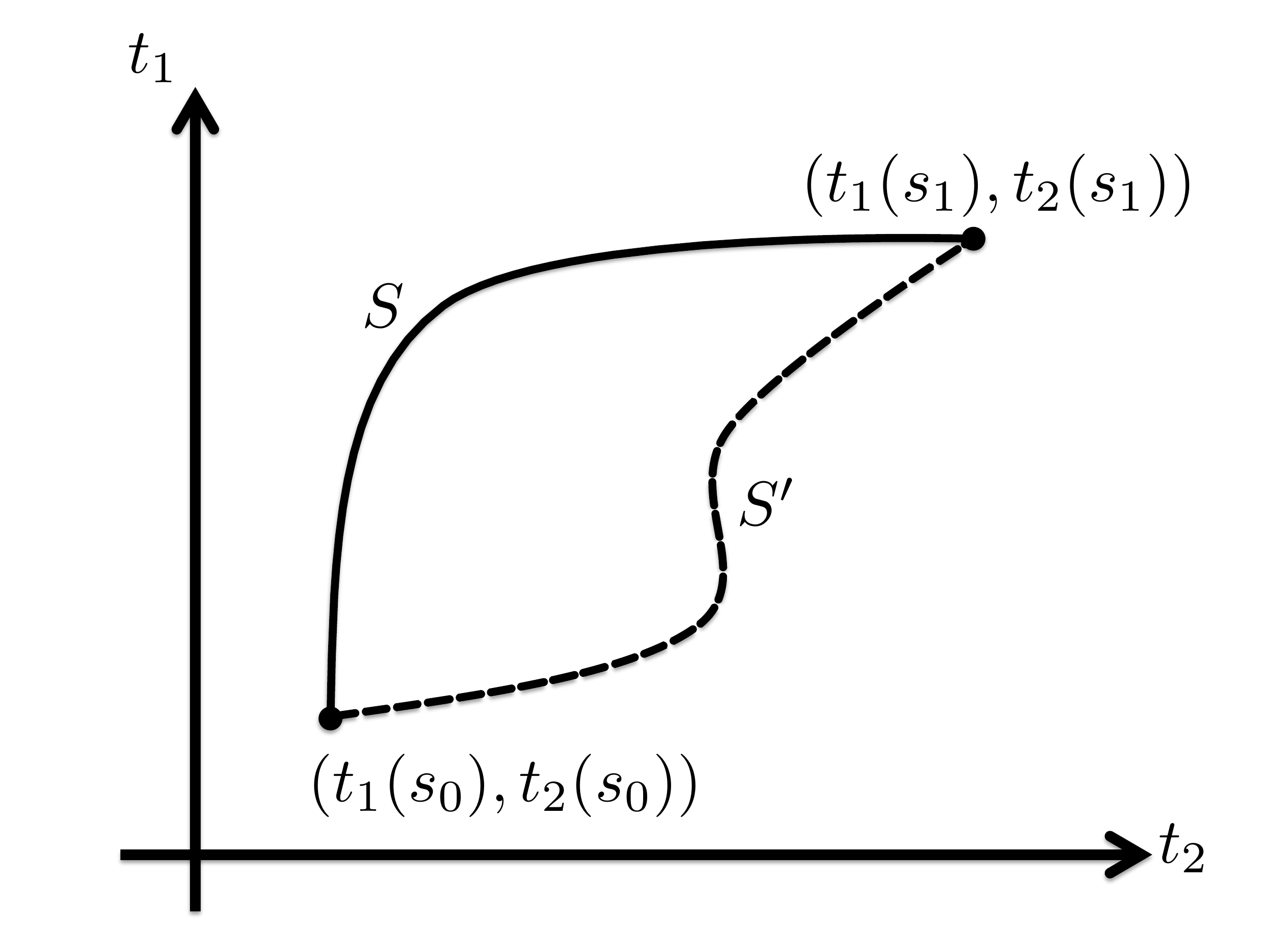} 
   \caption{\small{The deformation of the continuous curve on the space of the independent variables $(t_1(s),t_2(s))$, where $s$ is the time paramentised variable: $s_0<s<s_1$. The invariance of the action comes from the fact that $\delta S=S^\prime-S=0$, resulting in the closure relation. }} 
   \label{lattice22}
\end{figure}
Furthermore, we find that the closure relation for these two Lagrangians reads
\begin{equation}\label{conclosure}
\frac{\partial L_{(t_2)}}{\partial t_1}=\frac{\partial L_{(t_1)}}{\partial t_2}\;.
\end{equation}
Again, this relation guarantees the invariance of the action
\begin{equation}
S=\int_{\Gamma}\left(L_{(t_1)}dt_1+L_{(t_2)}dt_2\right)\;,
\end{equation}
under local deformation of the curve $\Gamma$ on the space of the independent variables $(t_1,t_2)$, see figure \ref{lattice22}.
\\
\\
\textbf{\emph{Remark}}: We find that the Lagrangians $L_{(t_1)}$ and $L_{(t_2)}$ have the same momentum variable
\begin{equation}
\pi_i=\ln\left|\frac{\partial X_i}{\partial t_1}\right|+1+\sum_{j=1,j\ne i}^N\ln(X_i-X_j)\;,\label{pi2}
\end{equation}
which can also be derived from the continuum limits of \eqref{pi}
\begin{eqnarray}
&&\pi_i=\sum_{j=1}^N\ln(x_j-\wh x_i)+1+\ln|q|,\nn\\
\xmapsto{\mbox{skew limit}}&&\pi_i=\sum_{j=1}^N\ln(\mathsf x_j-\bar {\mathsf{x}}_i)+1+\ln|p|,\nn\\
\xmapsto{\mbox{full limit}}&&\pi_i=\ln\left|\frac{\partial X_i}{\partial t_1}\right|+1+\sum_{j=1,j\ne i}^N\ln(X_i-X_j)\;.\nn
\end{eqnarray}
Here only the dominant terms are considered. 
\\
\\
Using \eqref{pi2}, we find that 
\begin{equation}
h_i^2=\frac{\partial X_i}{\partial t_1}=\frac{e^{\pi_i-1}}{\prod_{j\ne i}(X_i-X_j)}\;,
\end{equation}
and the $\boldsymbol{L}_0$ becomes
\begin{equation}
\boldsymbol{L}_0=\sum_{i,j=1}^Nh_ih_jE_{ij}=\sum_{ij=1}^N\sqrt{\frac{\partial X_i}{\partial t_1}\frac{\partial X_j}{\partial t_1}}E_{ij}\;,
\end{equation}
and 
\begin{equation}
\mbox{Tr}\boldsymbol{L}_0=H_{(t_1)}=\sum_{i=1}^N\frac{\partial X_i}{\partial t_1}=\sum_{i=1}^N\frac{e^{\pi_i-1}}{\prod_{j\ne i}(X_i-X_j)}\;,
\end{equation}
which is the first Hamiltonian in the hierarchy. The connection to the Lagrangian $L_{(t_1)}$ can be seen from Legendre transformation
\begin{eqnarray}
L_{(t_1)}&=&\sum_{i=1}^N\pi_i\frac{\partial X_i}{\partial t_1}-H_{(t_1)}
=\sum_{i=1}^N(\pi_i-1)\frac{\partial X_i}{\partial t_1}\nn\\
&=&\sum\limits_{i=1}^N\frac{\partial X_i}{\partial t_1}\ln\left| \frac{\partial X_i}{\partial t_1}\right|+\sum\limits_{i \ne j}^N \frac{\partial X_j}{\partial t_1}\ln|X_i-X_j|\nn\;,
\end{eqnarray}
with the help of \eqref{pi2}.
\\
\\
Unfortunately, the Lax matrix $\boldsymbol{L}_0$ has to be treated as a \emph{fake Lax matrix} since it produces only the first conserved quantity of motion \cite{Calogero goldfish}. To obtain the rest of the Hamiltonians, we need another method \cite{Suris}. Let us consider the second Hamiltonian given by
\[
H_{(t_2)}=\sum_{i=1}^N\frac{e^{\pi_i-1}}{\prod_{j\ne i}(X_i-X_j)}\sum_{j=1,j \ne i}^NX_j\;,
\]
and
\begin{equation}
\frac{\partial X_i}{\partial t_2}=\frac{\partial H_{(t_2)}}{\partial \pi_i}=\frac{e^{\pi_i-1}}{\prod_{j\ne i}(X_i-X_j)}\sum_{j=1,j \ne i}^NX_j\;.
\end{equation}
Performing the Legendre transformation
\begin{eqnarray}
L_{(t_2)}&=&\sum_{i=1}^N\pi_i\frac{\partial X_i}{\partial t_2}-H_{(t_2)}
=\sum_{i=1}^N(\pi_i-1)\frac{\partial X_i}{\partial t_2}\nn\\
&=&\sum\limits_{i=1}^N\frac{\partial X_i}{\partial t_2}\ln\left| \frac{\partial X_i}{\partial t_1}\right|+\sum\limits_{i \ne j}^N \frac{\partial X_j}{\partial t_1}\ln|X_i-X_j|\nn\;,
\end{eqnarray}
which is the second Lagrangian (up to the total derivative term).

\section{The connection to the lattice KP systems}\label{conKP}
In \cite{Sikarin1}, the discrete-time Calogero-Moser system was naturally obtained by looking at the pole-solution of the semi-discrete KP equation. In contrast, the discrete-time Ruijsenaars-Schneider system was constructed from Ansatz Lax pair. However, in \cite {Sikarin2}, the connect between the Ruijseenaars-Schneider system and the lattice KP systems was established. In the same fashion with the Ruijseenaars-Schneider system, we start to derive the discrete-time Calogero's goldfish from the Ansatz Lax pair. In this section, we will investigate the connection between the lattice KP systems and the Calogero's goldfish.
\\
\\
We start to consider the $\tau$-function as its characteristic polynomial: 
\begin{equation}\label{Sec}
\tau (\xi )=\det(\xi \boldsymbol I-\boldsymbol Y) \;,
\end{equation}
$\bsY=\bsY(n,m,h)$, given in \eqref{Ynm1}, is the function of three discrete variables and there are the relations
\begin{subequations}\label{KPYrels} 
\begin{eqnarray}\label{KPYn1}
\wt{\boldsymbol{Y}}-\boldsymbol{Y}&=&\wt{\boldsymbol{r}}\boldsymbol{s}^T\;,\\
\wh{\boldsymbol{Y}}-\boldsymbol{Y}&=&\wh{\boldsymbol{r}}\boldsymbol{s}^T\;,\\
\ol{\boldsymbol{Y}}-\boldsymbol{Y}&=&\ol{\boldsymbol{r}}\boldsymbol{s}^T\;,
\end{eqnarray}
\end{subequations}
where $\boldsymbol{r}$ and $\boldsymbol{s}$ are the functions of discrete variables via the following 
shift relations (see \eqref{rs}): 
\begin{subequations}\label{KPrs}\begin{eqnarray}
&& (p\boldsymbol{I}+\bLam)\cdot \wt{\boldsymbol r}=\boldsymbol r\;,\;\;\;\;\boldsymbol{s}^T\cdot (p\boldsymbol{I}+\bLam)=\wt{\boldsymbol s}^T\;, \\ 
&& (q\boldsymbol{I}+\bLam)\cdot \wh{\boldsymbol r}=\boldsymbol r\;,\;\;\;\;\boldsymbol{s}^T\cdot (q\boldsymbol{I}+\bLam)=\wh{\boldsymbol s}^T\;, \\ 
&& (r\boldsymbol{I}+\bLam)\cdot \ol{\boldsymbol r}=\boldsymbol r\;,\;\;\;\;\boldsymbol{s}^T\cdot (r\boldsymbol{I}+\bLam)=\ol{\boldsymbol s}^T\;.
\end{eqnarray}\end{subequations} 
To derive the lattice KP equations, we first perform the computation
\begin{eqnarray}
\wt{\tau}(\xi)&=&\det(\xi-\boldsymbol Y-\wt{\boldsymbol r}\boldsymbol s^T)\;,\nn\\
&=&\det((\xi-\boldsymbol Y)(1-\wt{\boldsymbol r}\boldsymbol s^T(\xi-\boldsymbol Y)^{-1}))\;,\nn\\
&=&\tau(\xi)(1-\boldsymbol s^T(\xi-\boldsymbol Y)^{-1}\wt{\boldsymbol r})\;,\nn
\end{eqnarray}
then we have
\begin{equation}\label{tau1}
\frac{\wt{\tau}(\xi)}{\tau(\xi)}=\boldsymbol {\mathrm v}_{p}(\xi)\;,
\end{equation}
in which the function $\boldsymbol {\mathrm{v}}_p$ is given by 
\begin{equation}\label{va}
\boldsymbol {\mathrm{v}}_a(\xi):= 1-\boldsymbol s^T(\xi-\boldsymbol Y)^{-1}(a+\bLam)^{-1}{\boldsymbol r}
\end{equation}
for a general parameter $a$. 
\\
\\
The reverse relation of Eq. \eqref{tau1} can be obtained by a similar computation: 
\begin{eqnarray}
\tau(\xi)&=&\det(\xi-\wt{\boldsymbol Y}+\wt{\boldsymbol r}\boldsymbol s^T)\;,\nn\\
&=&\det((\xi-\wt{\boldsymbol Y})(1+\wt{\boldsymbol r}\boldsymbol s^T(\xi-\wt{\boldsymbol Y})^{-1}))\;,\nn\\
&=&\wt{\tau}(\xi)(1+ {\boldsymbol s}^T(\xi-\wt{\boldsymbol Y})^{-1}\wt{\boldsymbol r})\;,\nn
\end{eqnarray}
then we have
\begin{equation}\label{tau2}
\frac{\tau(\xi)}{\wt{\tau}(\xi)}=\wt{\boldsymbol {\mathrm w}}_{p}(\xi)\;, 
\end{equation}
in which the function $\boldsymbol {\mathrm{w}}_p$ is given by 
\begin{equation}\label{wa}
\boldsymbol {\mathrm{w}}_a(\xi):= 1+\boldsymbol s^T(a+\bLam)^{-1}(\xi-\boldsymbol Y)^{-1}{\boldsymbol r},
\end{equation}
for a general parameter $a$.
\\
\\
From \eqref{tau1} and \eqref{tau2}, we have the relation
\begin{equation}\label{tauvw}
\frac{\tau(\xi)}{\wt{\tau}(\xi)}=\wt{\boldsymbol {\mathrm w}}_{p}(\xi)=\frac{1}{\boldsymbol {\mathrm v}_{p}(\xi)}\;.
\end{equation}
The same types of the relations for the other discrete directions can be obtained through the same computation
\begin{subequations}
\begin{eqnarray}\label{tauvw12}
&&\frac{\tau(\xi)}{\wh{\tau}(\xi  )}=\wh{\boldsymbol {\mathrm w}}_{q}(\xi )=\frac{1}{\boldsymbol {\mathrm v}_{q}(\xi)}\;,\\
&&\frac{\tau(\xi)}{\ol{\tau}(\xi   )}=\ol{\boldsymbol {\mathrm w}}_{r}(\xi   )=\frac{1}{\boldsymbol {\mathrm v}_{r}(\xi)}\;.
\end{eqnarray}
\end{subequations}
%
We now introduce the $N$-component vectors
 \begin{subequations}
\begin{eqnarray}
\boldsymbol{ u}_a(\xi)&=&(\xi-\boldsymbol Y)^{-1}(a+\bLam)^{-1}\boldsymbol{r}\;,\label{uu1}\\
\boldsymbol{\tbu}_b(\xi)&=&\boldsymbol{s}^T(b+\bLam)^{-1}(\xi-\boldsymbol Y)^{-1}\;,\label{uu2}
\end{eqnarray}
as well as the scalar variables
\begin{equation}\label{S}
S_{ab}(\xi)=\boldsymbol{s}^T(b+\bLam)^{-1}(\xi-\boldsymbol Y)^{-1}(a+\bLam)^{-1}\boldsymbol{r}\;.
\end{equation}
\end{subequations}
Equation \eqref{uu1} can be written in the form of
\begin{eqnarray}
\boldsymbol{ u}_a(\xi)&=&(p-a)\wt{\boldsymbol{ u}}_a(\xi )+\boldsymbol{\mathrm v}_a(\xi)\wt{\boldsymbol{ u}}_0(\xi )\;,\label{ua}
\end{eqnarray}
with $\boldsymbol{ u}_0(\xi)=(\xi-\boldsymbol Y)^{-1}\boldsymbol{r}$,
and equation \eqref{uu2} can also be rewritten as
\begin{equation}\label{ub}
\wt{\boldsymbol{ \tbu}}_b(\xi)=(p-b)\wt{\boldsymbol{ \tbu}}_b(\xi )+\wt{\boldsymbol{\mathrm w}}_b(\xi)\wt{\boldsymbol{ \tbu}}_0(\xi )\;,
\end{equation}
with $\boldsymbol{ \tbu}_0(\xi)=\boldsymbol{s}^T(\xi-\boldsymbol Y)^{-1}$.
\\
\\
Another type of relation can be obtained by multiply $\wt{\boldsymbol{s}}^T(b+\bLam)^{-1}$ on the left hand side of \eqref{ua}. We have
\begin{eqnarray}
\wt{\boldsymbol{s}}^T(b+\bLam)^{-1}\boldsymbol{ u}_a(\xi)&=&(p-a)\wt{\boldsymbol{s}}^T(b+\bLam)^{-1}\wt{\boldsymbol{ u}}_a(\xi )\nn\\
&&+\boldsymbol{\mathrm v}_a(\xi)\wt{\boldsymbol{s}}^T(b+\bLam)^{-1}\wt{\boldsymbol{ u}}_0(\xi
)\;,\nn\\
\boldsymbol{s}^T(p+\bLam)(b+\bLam)^{-1}\boldsymbol{ u}_a(\xi)&=&(p-a)\wt{S}_{ab}(\xi )+\boldsymbol{\mathrm v}_a(\xi)\wt{\boldsymbol{ \mathrm{w}}}_b(\xi )\;,\nn\\
\boldsymbol{\mathrm v}_a(\xi)\wt{\boldsymbol{ \mathrm{w}}}_b(\xi )&=&1+(p-b)S_{ab}(\xi)-(p-a)\wt{S}_{ab}(\xi )\;.\label{vw1}
\end{eqnarray}
%
%
%
Immediately, the other equations in other discrete-time directions are
 \begin{subequations}
\begin{eqnarray}
\boldsymbol{\mathrm v}_a(\xi)\wh{\boldsymbol{ \mathrm{w}}}_b(\xi  )&=&1+(q-b)S_{ab}(\xi)-(q-a)\wh{S}_{ab}(\xi  )\;,
\label{vw3}\\
\boldsymbol{\mathrm v}_a(\xi)\ol{\boldsymbol{ \mathrm{w}}}_b(\xi  )&=&1+(r-b)S_{ab}(\xi)-(r-a)\ol{S}_{ab}(\xi  )\;.
\label{vw3}
\end{eqnarray}
\end{subequations}
Using the identity
\begin{equation}
\frac{\wt{\overline{\boldsymbol{\mathrm w}}}_b(\xi )\overline{\boldsymbol{ \mathrm{v}}}_a(\xi  )}{\wh{\overline{\boldsymbol{\mathrm w}}}_b(\xi  )\overline{\boldsymbol{ \mathrm{v}}}_a(\xi  )}
=\frac{\wt{\overline{\boldsymbol{\mathrm w}}}_b(\xi  )\wt{\boldsymbol{ \mathrm{v}}}_a(\xi   )}{\wh{\overline{\boldsymbol{\mathrm w}}}_b(\xi )\wh{\boldsymbol{ \mathrm{v}}}_a(\xi   )}\;
\frac{\wh{\wt{\boldsymbol{\mathrm w}}}_b(\xi  )\wh{\boldsymbol{ \mathrm{v}}}_a(\xi )}{\wh{\wt{\boldsymbol{\mathrm w}}}_b(\xi )\wt{\boldsymbol{ \mathrm{v}}}_a(\xi  )}\;,
\label{idenvw}
\end{equation}
we can derive
\begin{eqnarray}\label{SKP}
&&\frac{1+(p-b)\overline{S}_{ab}(\xi  )-(p-a)\wt{\overline S}_{ab}(\xi  )}{1+(q-b)\overline{S}_{ab}(\xi   )-(q-a)\wh{\overline S}_{ab}(\xi )}\nn\\
&&\;\;=\frac{1+(r-b)\wt S_{ab}(\xi   )-(r-a)\wt{\overline{S}}_{ab}(\xi )}{1+(q-b)\wt S_{ab}(\xi   )-(q-a)\wh{\wt{S}}_{ab}(\xi  )}
 \frac{1+(p-b)\wh S_{ab}(\xi  )-(p-a)\wh{\wt{S}}_{ab}(\xi  )}{1+(r-b)\wh S_{ab}(\xi   )-(r-a)\wh{\overline{S}}_{ab}(\xi  )}\;,\nn\\
\end{eqnarray}
which is a three-dimensional lattice equation which first appeared  in 
\cite{NCWQ}, or the \emph{Schwarzian lattice KP equation} \cite{DN}.
\\
\\
We now multiply $\wt{\boldsymbol s}^T$ on the left hand side of \eqref{ua} leading to
\begin{eqnarray}\label{us}
\wt{\boldsymbol s}^T\boldsymbol{ u}_a(\xi)&=&(p-a)\wt{\boldsymbol s}^T\wt{\boldsymbol{ u}}_a(\xi )+\boldsymbol{\mathrm v}_a(\xi)\wt{\boldsymbol s}^T\wt{\boldsymbol{ u}}_0(\xi )\;,\nn\\
\boldsymbol s^T(p+\bLam)\boldsymbol{ u}_a(\xi)&=&(p-a)(1-\wt{\boldsymbol{ \mathrm{v}}}_a(\xi ))+\boldsymbol{\mathrm v}_a(\xi)\wt{\boldsymbol s}^T\wt{\boldsymbol{ u}}_0(\xi )\;.
\end{eqnarray}
Introducing 
\begin{equation}\label{u00}
u(\xi)=\boldsymbol s^T(\xi-\boldsymbol Y)^{-1}\boldsymbol r\;,
\end{equation}
Equation \eqref{us} can be written in the form
\begin{eqnarray}\label{u001}
(p+\wt{u}(\xi ))\boldsymbol{\mathrm v}_a(\xi)-(p-a)\wt{\boldsymbol{\mathrm v}}_a(\xi)=a+\boldsymbol s^T\bLam\boldsymbol{ u}_a(\xi)\;.
\end{eqnarray}
Another two relations related to the other discrete directions can be automatically obtained
\begin{subequations}
\begin{eqnarray}
(q+\wh{u}(\xi  ))\boldsymbol{\mathrm v}_a(\xi)-(q-a)\wh{\boldsymbol{\mathrm v}}_a(\xi  )&=&a+\boldsymbol s^T\bLam\boldsymbol{ u}_a(\xi)\label{u001}\;,\\
(r+\overline{u}(\xi  ))\boldsymbol{\mathrm v}_a(\xi)-(r-a)\overline{\boldsymbol{\mathrm v}}_a(\xi  )&=&a+\boldsymbol s^T\bLam\boldsymbol{ u}_a(\xi)\label{u001}\;.
\end{eqnarray}
\end{subequations}
Eliminating the term $\boldsymbol s^T\bLam\boldsymbol{ u}_a(\xi)$, we can derive the relations
\begin{subequations}
\begin{eqnarray}\label{set1}
(p-q+\wt{u}(\xi )-\wh{u}(\xi  ))\boldsymbol{\mathrm v}_a(\xi)&=&(p-a)\wt{\boldsymbol{\mathrm v}}_a(\xi )
-(q-a)\wh{\boldsymbol{\mathrm v}}_a(\xi  )\label{u00vw1}\;,\\
(p-r+\wt{u}(\xi )-\overline{u}(\xi  ))\boldsymbol{\mathrm v}_a(\xi)&=&(p-a)\wt{\boldsymbol{\mathrm v}}_a(\xi )
-(r-a)\overline{\boldsymbol{\mathrm v}}_a(\xi  )\label{u00vw2}\;,\\
(r-q+\overline{u}(\xi  )-\wh{u}(\xi  ))\boldsymbol{\mathrm v}_a(\xi)&=&(r-a)\overline{\boldsymbol{\mathrm v}}_a(\xi  )
-(q-a)\wh{\boldsymbol{\mathrm v}}_a(\xi  )\label{u00vw3}\;.
\end{eqnarray}
\end{subequations}
We now set $p=a$ then \eqref{u00vw1} and \eqref{u00vw2} become
\begin{subequations}\label{set2}
\begin{eqnarray}
p-q+\wt{u}(\xi )-\wh{u}(\xi  )&=&-(q-p)\frac{\wh{\boldsymbol{\mathrm v}}_p(\xi    )}{\boldsymbol{\mathrm v}_p(\xi)}\label{u00vw11}\;,\\
p-r+\wt{u}(\xi )-\overline{u}(\xi )&=&-(r-p)\frac{\overline{\boldsymbol{\mathrm v}}_p(\xi    )}{\boldsymbol{\mathrm v}_p(\xi)}\label{u00vw21}\;,
\end{eqnarray}
\end{subequations}
The combination of \eqref{u00vw11} and \eqref{u00vw21} gives
\begin{equation}\label{LKP}
\frac{p-q+\wt{u}(\xi )-\wh{u}(\xi  )}{p-r+\wt{u}(\xi )-\overline{u}(\xi  )}
=\frac{p-q+\wt{\overline u}(\xi  )-\wh{\overline u}(\xi  )}{p-r+\wh{\wt{u}}(\xi  )-\wh{\overline{u}}(\xi   )}\;,
\end{equation}
which is the ``\emph{lattice KP equation}" \cite{NCWQ}, cf. also \cite{NCW}. 
\\
\\
From the definition of the function $\boldsymbol{\mathrm v}_p(\xi)$ in \eqref{tau1}, \eqref{u00vw11} and \eqref{u00vw21} can be written in terms 
of the $\tau$-function
\begin{subequations}
\begin{eqnarray}
p-q+\wt{u}(\xi )-\wh{u}(\xi  )&=&-(q-p)\frac{\wh{\wt{\tau}}(\xi )}{\wh{\tau}(\xi  )}\frac{\tau(\xi)}{\wt{\tau}(\xi )}\label{u00vw113}\;,\\
p-r+\wt{u}(\xi )-\overline{u}(\xi  )&=&-(r-p)\frac{\wt{\overline{\tau}}(\xi )}{\overline{\tau}(\xi  )}\frac{\tau(\xi)}{\wt{\tau}(\xi )}\label{u00vw213}\;.
\end{eqnarray}
\end{subequations}
From \eqref{u00vw3}, if we set $r=a$ we also have
\begin{eqnarray}
r-q+\overline{u}(\xi )-\wh{u}(\xi  )=-(q-r)\frac{\wh{\overline{\tau}}(\xi  )}{\wh{\tau}(\xi  )}\frac{\tau(\xi)}{\overline{\tau}(\xi  )}\label{u00vw114}\;.
\end{eqnarray}
The combination of \eqref{u00vw113} \eqref{u00vw213} \eqref{u00vw114} yields
\begin{eqnarray}\label{Hirota}
&&(p-q)\wh{\wt{\tau}}(\xi  )\overline{\tau}(\xi  )+(r-p)\wt{\overline{\tau}}(\xi  )\wh{\tau}(\xi )+(r-q)\wh{\overline{\tau}}(\xi  )\wt{\tau}(\xi )=0\;,
\end{eqnarray}
which is the bilinear lattice KP equation, (originally coined DAGTE, cf. \cite{Hirota}).
\\
\\
We managed to establish the connection between the Calogero's goldfish system and the lattice KP systems. This completes the picture of the connection between discrete integrable one dimensional many-body systems, namely Calogero-Moser system, Ruijsenaars-Schneider system and Calogero's goldfish system, with the lattice KP systems.

\section{Summary}
Another concrete example for the Lagrangian 1-form was studied through the rational Calogero's goldfish system in full detail. In this example, at the discrete-time level, the system was obtained from the Ansatz Lax pair, rather through the pole-reduction process of the KP system in discrete-time Calogero-Moser system, like those for the case of discrete-time Ruijsenaars-Schneider system leading to a system of discrete-time Calogero's goldfish systems associated with different discrete variables. The compatibility between these two discrete direction provided the constraints telling how the system moves from one discrete variable to another discrete variable. The variation of the discrete action with respect to discrete-time variable resulting the closure relation which guarantees the unchanged value of the action under local deformation of the discrete curve on the space of discrete-time variables. Then the continuum limits had been applied to the system, namely the skew limit and the full continuum limit, in order to generate the Lagrangian hierarchy of the system. Intriguingly, these Lagrangians are the function of many-time variables (the number of time variables is up to the number of the particles in this case). The continuous closure relation of the system, resulting directly from the variational principle with respect to time variables, again guarantees the invariant of the action under the local deformation of the continuous curve on the space of continuous variables. Furthermore, the connection between the Calogero's goldfish system and the lattice KP systems was established through the structure of the exact solution of the Calogero's goldfish system.


\begin{acknowledgements}
Sikarin Yoo-Kong gratefully acknowledges the support from the Thailand Research Fund (TRF) under grant number: TRG5680081.
\end{acknowledgements}


\begin{thebibliography}{}
%

\bibitem{SF1} Lobb S B and Nijhoff F W 2009,  \emph{Lagrangian multiforms and multidimensional consistency}, J. Phys. A: Math. Theor. \textbf{42}, 454013.

\bibitem{SF2} Lobb S B Nijhoff F W 2010,  \emph{Lagrangian multiform structure for the lattice Gel' fand-Dikii hierarchy}, J. Phys. A: Math. Theor. \textbf{43}, 072003.

\bibitem{SF3} Lobb S B, Nijhoff F W and Quispel G R W 2009,  \emph{Lagrangian multiforms structure for the lattice KP system}, J. Phys. A: Math. Theor. \textbf{42}, 472002.

\bibitem{Original CM1} Calogero F 1969,  \emph{Solution of a three-body problem in one dimension}, J. Math. Phys. \textbf{42}, pp.2191-2196.

\bibitem{Original CM2} Calogero F 1971,  \emph{Solution of the one-dimensional
N-body problem with quadratic and/or inversely quadratic pair
potentials}, J. Math. Phys. \textbf{12}, pp.418-436.

\bibitem{Sikarin1}  Yoo-Kong S, Lobb S and Nijhoff F 2011, \emph{Discrete-time Calogero-Moser system and Lagrangian 1-form structure}, J. Phys. A: Math. Theor, \textbf{44}, 365203.

\bibitem{Sikarin2}  Yoo-Kong S and Nijhoff F 2013, \emph{Discrete-time Ruijsenaars-Schneider system and Lagrangian 1-form structure}, ArXiv:1112.4576v2 {\tt nlin.SI }.

\bibitem{Toda}  Boll R, Petrera M and Yuri B S 2013, \emph{Multi-time Lagrangian 1-forms for families of Backlund transformations. Toda-type systems}, J. Phys. A:Math. Theor. \textbf{46}, 275204.

\bibitem{Toda2}  Boll R, Petrera M and Yuri B S 2014, \emph{Multi-time Lagrangian 1-forms for families of Backlund transformations. Relativistic Toda-type systems}, arXiv:1408.2405.

\bibitem{Calogero goldfish}  Calogero F 2001, \emph{The neatest many-body problem amenable to exact treatments (a ``goldfish")}, Phys. D, pp.78-84. 

\bibitem{Suris}  Yuri B S 2005, \emph{Time Discretization of F. Calogero's ``Goldfish" System}, J. Non. Math. Phys. \textbf{12}, pp.633-647.

\bibitem{NCWQ}
Nijhoff F W, Capel H W, Wiersma G L, and Quispel G R W 1984, 
\emph{Bucklund transformations and three-dimensional lattice equations}, 
Phys. Lett. A, \textbf{105}, pp.267-272.

\bibitem{DN}
Dorfman I Ya, Nijhoff F W 1991, 
\emph{On a (2+1)-dimensional version of the Krichever-Novikov equation}, 
Phys. Lett. A, \textbf{157}, pp.107-112.

\bibitem{NCW}
Nijhoff F W, Capel H W, and Wiersma G L 1985, 
\emph{Integrable lattice systems in two and
three dimensions}, In: Geometric Aspects of the Einstein Equations and Integrable
Systems, Ed. R. Martini, Lecture Notes in Physics, Berlin/New York, Springer Verlag, pp.263-302.

\bibitem{Hirota} Hirota R 1981,
\emph{Discrete Anologue of a Generalised Toda Equation}, 
J. Phys. Soc. Japan, \textbf{50}, pp.3785-3791.
\bibitem{Niff} Nijhoff F W and Pang G D 1996, \emph{Discrete-time Calogero-Moser model
and lattice KP equations}, ArXiv:9409071.




\end{thebibliography}
\end{document}